# Emulated Inertia and Damping of Converter-Interfaced Power Source

Bin Wang, *Student Member, IEEE*, Yichen Zhang, *Student Member, IEEE*, Kai Sun, *Senior Member, IEEE*, Kevin Tomsovic, *Fellow, IEEE*

*Abstract*—Converter-interfaced power sources (CIPSs), like wind turbine and energy storage, can be switched to the inertia emulation mode when the detected frequency deviation exceeds a pre-designed threshold, i.e. dead band, to support the frequency response of a power grid. This letter proposes an approach to derive the emulated inertia and damping from a CIPS based on the linearized model of the CIPS and the power grid, where the grid is represented by an equivalent single machine. The emulated inertia and damping can be explicitly expressed in time and turn out to be time-dependent.

## I. INTRODUCTION

INCREASING renewables constantly integrated into a power grid keeps decreasing the system inertia, which may lead to a worse frequency nadir after, e.g., a loss of generation. It is reported that converter-interfaced power sources (CIPSs), e.g. wind turbine and energy storage, have the capability to provide additional power by switching in an additional control loop when detecting a large enough frequency deviation [1]-[4]. Such capability is also desired by the industry [5]. However, although many papers have studied the capacity of support from CIPS, their inertia and damping seen from the system side have not been well defined during a disturbance, which requires the consideration of the additional control loop. This letter proposed a way to formulate the emulated inertia, also called synthetic inertia, and emulated damping of wind turbine or energy storage as explicit functions of time, which may enable dynamic analyses considering the change in parameters like inertia and damping.

## II. EMULATED INERTIA AND DAMPING OF CIPS

Suppose the power grid can be equivalent to a single machine. The reference value of the active power for the control of CIPS under large frequency deviation are usually designed by (1), where $P_{ref}$ represents the reference active power during normal condition, $K_{dr}$ and $K_{ie}$ are coefficients of the additional control loop respectively emulating damping and inertia effects.

$$P_{total,ref} = P_{ref} + K_{dr}\Delta\omega + K_{ie}\Delta\dot{\omega} \qquad (1)$$

B. Wang, Y. Zhang, K. Sun and K. Tomsovic are with the University of Tennessee, Knoxville, TN 37996 USA (e-mail: bwang@utk.edu; yzhan124@vols.utk.edu; kaisun@utk.edu; tomsovic@eecs.utk.edu).

The general structure of a CIPS connected to the power grid is shown in Fig.1. When the CIPS is switched to the inertia/damping emulation mode, the frequency deviation (FD) and the rate of change of frequency (ROCOF) from the grid side have to be involved in the dynamics of the CIPS, as seen in (1), which can be treated as the inputs of the CIPS. The underlying equations are shown in (2) and (3).

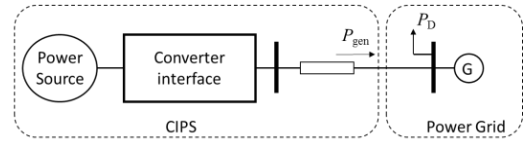

Fig.1. Structure of the CIPS connected to an equivalent single machine

$$\text{CIPS}: \begin{cases} \begin{bmatrix} \dot{x} \\ 0 \end{bmatrix} = \begin{bmatrix} F(x,y,V,\Delta\omega,\Delta\dot{\omega}) \\ G(x,y,V,\Delta\omega,\Delta\dot{\omega}) \end{bmatrix} \\ P_{gen} = H(x,y,V,\Delta\omega,\Delta\dot{\omega}) \end{cases} \qquad (2)$$

$$\text{Grid}: \begin{cases} \Delta\dot{\omega} = \dfrac{\omega_s}{2H}\left(P_m + P_{gen} - P_D\right) - D\Delta\omega \\ \dot{z} = h(z) \\ V = g(x,y,z,\Delta\omega,\Delta\dot{\omega}) \end{cases} \qquad (3)$$

where $x$ and $y$ are the state vectors respectively containing all differential and algebraic variables of the CIPS; $P_{gen}$ represents the active power output of the CIPS; $V$ is the terminal bus voltage of the single machine; $\Delta\omega$ and $\Delta\dot{\omega}$ are respectively frequency deviation and its derivative of the equivalent single machine (to mimic the FD and ROCOF of the power grid); $z$ is the vector of all differential and algebraic variables of the single machine and the load except for $\Delta\dot{\omega}$ and $V$; $H$, $D$ and $P_m$ are respectively the inertia, damping and mechanical power of the single machine; and $P_D$ is the load.

Linearizing (2) at the stable equilibrium point $(x, y, z, V, \Delta\omega, \Delta\dot{\omega}) = (x_{ep}, y_{ep}, z_{ep}, V_{ep}, 0, 0)$, eliminating $y$ and $V$ and using variations for all variables give (4). Similarly, linearizing the first equation of (3) and using variations for all variables gives (5).

$$\begin{cases} \Delta\dot{x} = A\Delta x + B_1\Delta\omega + B_2\Delta\dot{\omega} \\ \Delta P_{gen} = C\Delta x + D_1\Delta\omega + D_2\Delta\dot{\omega} \end{cases} \qquad (4)$$

$$\Delta\dot{\omega} = \dfrac{\omega_s}{2H}\left(\Delta P_m + \Delta P_{gen} - \Delta P_D\right) - D\Delta\omega \qquad (5)$$

Then, the solution of the first equation in (4) with the initial condition $\Delta x(0)$ can be written in (6) according to superposition, where $\Delta x_0$, $\Delta x_1$ and $\Delta x_2$ respectively are the solutions of the three linear initial-value problems defined in (7). By linear control theory, the solutions to (7) are shown in (8).

$$\Delta x(t) = \Delta x_0(t) + \Delta x_1(t) + \Delta x_2(t) \tag{6}$$

$$\begin{cases} \Delta \dot{x}_0 = A\Delta x_0 & \text{with } \Delta x_0(0) = \Delta x(0) \\ \Delta \dot{x}_1 = A\Delta x_1 + B_1 \Delta \omega & \text{with } \Delta x_1(0) = 0 \\ \Delta \dot{x}_2 = A\Delta x_2 + B_2 \Delta \dot{\omega} & \text{with } \Delta x_2(0) = 0 \end{cases} \tag{7}$$

$$\begin{cases} \Delta x_0(t) = e^{At} x(0) \\ \Delta x_1(t) = \int_0^t e^{A(t-\tau)} B_1 \Delta \omega(\tau) d\tau \\ \Delta x_2(t) = \int_0^t e^{A(t-\tau)} B_2 \Delta \dot{\omega}(\tau) d\tau \end{cases} \tag{8}$$

Generally speaking, the last two integrals in (8) cannot be worked out into certain explicit forms since the unknown functions $\Delta \omega$ and $\Delta \dot{\omega}$. To overcome this hurdle, we will look for certain functions which are always capable to describe $\Delta \omega$ and $\Delta \dot{\omega}$, such that (i) the explicit solutions of the last two integrals in (8) exist and (ii) the second equation of (4) can be written into (9), where $a_0$, $a_1$ and $a_2$ do not explicitly depend on $\Delta \omega$ or $\Delta \dot{\omega}$. If such functions can be achieved, after substituting (9) into (5) to give (10), the emulated inertia and damping of the CIPS will be explicitly defined and follow the form of (11).

$$\Delta P_{gen} = a_0(t) + a_1(t)\Delta \omega + a_2(t)\Delta \dot{\omega} \tag{9}$$

$$\Delta \dot{\omega} = \frac{\omega_s}{2H + 2 \cdot \frac{\omega_s a_2(t)}{2}}(\Delta P_m - \Delta P_e) - \left(D - \frac{\omega_s a_1(t)}{2H}\right)\Delta \omega \tag{10}$$

$$\begin{cases} H_e(t) = \frac{\omega_s a_2(t)}{2} \\ D_e(t) = -\frac{\omega_s a_1(t)}{2H} \end{cases} \tag{11}$$

The following proposes a general form for functions of $\Delta \omega$ and $\Delta \dot{\omega}$ to achieve the above goal and then gives an example.

Typical responses of $\Delta \omega$ and $\Delta \dot{\omega}$ have the pattern shown in Fig.2. Thus, we assume that the general forms of $\Delta \omega$ and $\Delta \dot{\omega}$ respectively follow those shown in (12), where $P_n(t)$ is a polynomial function in $t$ up to a certain order $n$. Note that the integral in the second formula of (12) can always be worked out.

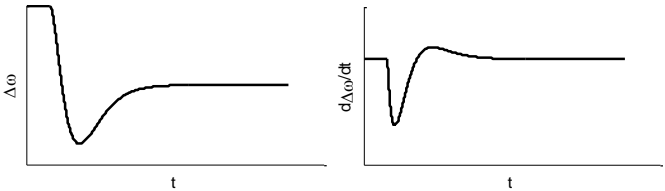

Fig.2. Typical responses of $\Delta \omega$ and $\Delta \dot{\omega}$ by (12) with $P_2(t) = -2.5t+t^2$ and $b=1$.

$$\begin{cases} \Delta \dot{\omega} = P_n(t)e^{-bt} = (c_0 + c_1 t^2 + \cdots + c_n t^n)e^{-bt} \\ \Delta \omega = \int_0^t P_n(\tau)e^{-b\tau} d\tau \end{cases} \tag{12}$$

For convenience, an example will be presented below with $P_1(t) = at$. With this assumption, we have $\Delta \omega$ and $\Delta \dot{\omega}$ as shown in (13). Then, the third equation in (8) can be rewritten as (14). Substitute (13) only into $\Delta \dot{\omega}$ in the integrand and denominator and then the integral can be worked out as (15) [6]. Similarly, we have (17) for the second integral in (8). Finally, (9) becomes (19) and the emulated inertia and damping are defined as (20).

$$\begin{cases} \Delta \dot{\omega} = ate^{-bt} \\ \Delta \omega = \frac{a - ae^{-bt} - abte^{-bt}}{b^2} \end{cases} \tag{13}$$

$$\Delta x_2(t) = \frac{\int_0^t e^{A(t-\tau)} B_2 \Delta \dot{\omega}(\tau) d\tau}{\Delta \dot{\omega}(t)} \cdot \Delta \dot{\omega}(t) \tag{14}$$

$$\Delta x_2(t) = \frac{\int_0^t e^{A(t-\tau)} B_2 a\tau e^{-b\tau} d\tau}{ate^{-bt}} \cdot \Delta \dot{\omega}(t) = \mu_2(t) \cdot \Delta \dot{\omega}(t) \tag{15}$$

$$\mu_2(t) = \frac{-B_2 e^{-bt}(A+bI)^{-1}\left(t+(A+bI)^{-1}\left(I - e^{(A+bI)t}\right)\right)}{te^{-bt}} \tag{16}$$

$$\Delta x_1(t) = \frac{\int_0^t e^{A(t-\tau)} B_1 \cdot \frac{a - ae^{-b\tau} - ab\tau e^{-b\tau}}{b^2} \cdot d\tau}{\frac{a - ae^{-bt} - abte^{-bt}}{b^2}} \cdot \Delta \omega(t) = \mu_1(t) \cdot \Delta \omega(t) \tag{17}$$

$$\mu_1(t) = \left\{-B_1 A^{-1}\left(I - e^{At}\right) + B_1 e^{-bt}(A+bI)^{-1}\left(I - e^{(A+bI)t}\right) \right. \\ \left. + bB_1 e^{-bt}(A+bI)^{-1}\left(t+(A+bI)^{-1}\left(I - e^{(A+bI)t}\right)\right)\right\} / \left(1 - e^{-bt} - bte^{-bt}\right) \tag{18}$$

$$\Delta P_{gen} = a_0(t) + (D_1 + C\mu_1(t))\Delta \omega + (D_2 + C\mu_2(t))\Delta \dot{\omega} \tag{19}$$

$$\begin{cases} H_e(t) = \frac{\omega_s}{2}(D_2 + C\mu_2(t)) \\ D_e(t) = -\frac{\omega_s}{2H}(D_1 + C\mu_1(t)) \end{cases} \tag{20}$$

REFERENCES

[1] F. D. Gonzalez, A. Sumper, O. G. Bellmunt, *Energy Storage in Power Systems*, Hoboken: Wiley, 2016.
[2] J. Morren, S. W. H. de Haan, W. L. Kling, J. A. Ferreira, "Wind turbines emulating inertia and supporting primary frequency control," *IEEE Trans. Power Syst.*, vol.21, no.1, pp.433-434, Feb. 2006.
[3] H Ye, W Pei, Z. Qi, "Analytical modeling of inertial and droop responses from a wind farm for short-term frequency regulation in power systems," *IEEE Trans. Power Syst.*, vol.31, no.5, pp.3414-3423, Sep. 2016.
[4] F. D. Gonzalez, M. Hau, A. Sumper, O. G. Bellmunt, "Participation of wind power plants in system frequency control: review of grid code requirements and control methods," *Renew. Sustain. Energy Rev.*, vol.34, pp.551-564, 2014.
[5] J. Brisebois and N. Aubut, "Wind farm inertia emulation to fulfill Hydro-Québec's specific need," *IEEE PES General Meeting*, San Diego, 2011.
[6] D. Rowell, "Time-domain solution of LTI state equations," Class Handout in *Analysis and Design of Feedback Control System*, Oct. 2002.